\documentclass[letterpaper,journal]{IEEEtran}
\usepackage{amsmath,amsfonts}
\usepackage{algorithmic}
\usepackage{algorithm}
\usepackage{array}
\usepackage[caption=false,font=normalsize,labelfont=sf,textfont=sf]{subfig}
\usepackage{textcomp}
\usepackage{stfloats}
\usepackage{url}
\usepackage{verbatim}
\usepackage{graphicx}
\usepackage{cite}
\usepackage{hyperref}

\usepackage{todonotes}

\begin{document}
\title{Indicators of resilience for autonomous control systems}

\author{Jasper J. van Beers$^{1,2}$\thanks{$^{1}$These authors are with the Faculty of Aerospace Engineering, Delft University of Technology, 2629 HS Delft, The Netherlands}\thanks{$^{2}$Corresponding author: {\tt\small j.j.vanbeers@tudelft.nl}}, Da-Hwi Kim$^{1}$, Prashant Solanki$^{1}$, and Coen C. de Visser$^{1}$
}
\maketitle

\begin{abstract}
As modern societies rely more on autonomous systems to facilitate daily life, assuring their safe operation is paramount. Naturally, there are many techniques available to predict and prevent system failures. However, the safety afforded by such schemes may become misaligned with the true system, which can change in unexpected ways - from partial faults to natural wear-and-tear - that subtly degrade its stability. The implications that such subtle changes have on autonomous system stability can be observed through generic indicators of resilience derived from critical slowing down, popular for anticipating catastrophic tipping points in natural systems. Here, we show how one can systematically design these generic indicators for nonlinear control systems and show how these can reflect loss of stability though simulations of canonical robotic systems wherein their proximity to instability is manipulated directly. These results are affirmed through real-world flight experiments of a quadrotor that is nudged towards instability by progressively damaging its propeller blades. Our results show that the implications of degraded resilience on closed-loop stability are evident well before they appear, for which the indicators of resilience derived here can provide an early warning. 
\end{abstract}

\begin{IEEEkeywords}
autonomous systems, robot safety, nonlinear systems, critical slowing down 
\end{IEEEkeywords}

\section{Introduction}
Humanity depends on various autonomous control systems, from autopilots in self-driving cars and aircraft to telecommunication networks, yet these systems are not infallible. Subtle changes over their life cycle - such as partial faults and simple wear-and-tear - can work to degrade system resilience, challenging efforts to ensure their safe and sustained operation. These efforts would benefit from knowing when the system's resilience begins to deteriorate, which can help improve safety awareness, prevent failure and, ultimately, expedite the integration of various autonomous and robotic systems in society. 

An intuitive strategy to address this issue of safety is to directly estimate and accommodate the problems that a system encounters. For example, many have developed diverse designs for fault detection and fault-tolerant control that can identify expected failure modes and maintain system stability \cite{YU2015_FaultsAndSafety,Ke_2023_Quadrotor_123FTC}. Likewise, the field of robust control pursues designs that can guarantee safety and performance despite uncertainties \cite{zhou1996robust,TIJMEN2023_INDI_robustness}. Alternatively, safe operating spaces for a system can be derived based on its dynamic capabilities through reachability analysis \cite{mitchell2005time,LYGEROS2004917,bansal2017hamilton}. These safe spaces can then be used to develop early warning mechanisms, such as flight envelope protection systems on civil aircraft \cite{Beclastro2017_AC_LOCProblem,Schuet2017_AutonomousSFE_LOC}. However, these approaches share a common limitation in their dependence on system models to establish or monitor system safety. Indeed, many real world autonomous systems often suffer additional complexities and uncertainties that are challenging, if not impossible, to capture fully through modeling alone.

Recognizing this, recent advancements increasingly rely on data-driven schemes to help estimate safe operating spaces \cite{bansal2020deepreach,Zhang2019_OnlineSFEPrediction} or anticipate problematic modes of the system \cite{Panahi2026_UnsupervisedLearningTransitions,Ghadami_DataDrivenTippingPredictionSurvey}. In particular, machine learning is often leveraged to abstract the safe operating space from a family of representative models or measurement data \cite{NOROUZI2019434,Zhai2025_UnknownDynamics,Sun2019_MonteCarlo}. Likewise, any available data surrounding a destabilizing event can be used to design forecasters that anticipate the instability, such as for loss of control in quadrotors \cite{RNN_LOC_Altena}. However, many of these learning-based schemes require extensive data sets for learning, which are often lacking in the context of system instability and can be impractical to obtain. While such data reliance can be limited, it is often achieved by turning once again to models or system-specific insights \cite{Riso2020_flutter,Panahi2026_UnsupervisedLearningTransitions}. Such strategies may not generalize well to other destabilizing mechanisms or systems. 

The primary challenge with monitoring system resilience and instability appears to be that either model knowledge or a substantial amount of data is needed, but both are inherently lacking. This issue is compounded by the diversity of robotic systems, applications, and designs. Ideally, a system's resilience can be monitored without much intervention, modeling effort, and training data. Instead, it should be inferred directly from (real-time) measurement data and methods should be transferable across different systems.

Such requirements may be addressed by forecasting techniques developed for the natural world, where complexities challenge modeling and data is notoriously limited. For example, there is no fundamental barrier preventing the application of critical slowing down (CSD) - whereby a system's response rate slows as it approaches a bifurcation or `tipping point' \cite{Scheffer2009} - to monitor resilience in autonomous systems. Due to it's generic (i.e., system model-free) nature, CSD is widely used to predict critical transitions across a plethora of dynamic systems, ranging from anticipating collapse in the Earth's climate and ecosystems \cite{Scheffer2009,forzieri2022emerging,ditlevsen2023warning_AMOC_CSD}, to biological and epidemiological mechanisms \cite{Maturana2020,DeLeemput2014}, and also engineered systems \cite{kerr2023haptic,Pirani2024_NetworkCSD}. In particular, recent work has shown that CSD can also be expected in control systems, reflecting loss of stability in damaged quadrotors \cite{vanBeers2026_ews_for_loc}. While the underlying theory suggests that CSD applies generally to nonlinear control systems, it remains unclear how one should parameterize the CSD-based early warning signals for this. Moreover, the theoretical foundation behind the results of \cite{vanBeers2026_ews_for_loc} would benefit from a connection to dynamic models of the system, showing that the same trends can be expected in both simulated models and the real world.

The main contributions of this work are: (i) an overview of when CSD can be expected in nonlinear autonomous control systems and how it arises; (ii) using insights from control theory, we provide a systematic procedure to parameterize early warning signals for monitoring control system resilience; (iii) demonstrate that these resilience indicators reflect the proximity of a control system to closed-loop instability and resilience (i.e., robustness) through systematic simulations and real-world experiments of two canonical robotic systems: the inverted pendulum and the quadrotor.

Specifically, we use simulations of these systems to unambiguously drive them towards instability and show how CSD reflects this loss of stability, which can already be observed around relaxed operating modes such as hovering flight for a quadrotor. Furthermore, we show how these simulation results appear also in real world flight of damaged quadrotors when using the same resilience indicator parameterization, lending support to the theoretical foundations of the approach. We emphasize that these indicators of resilience are not intended as a replacement of existing fault diagnostic and safety schemes, rather they can be used as an add-on monitor providing an additional layer of safety for autonomous and robotic systems.

\section{Critical slowing down}\label{sec:csd}
Critical slowing down (CSD) predicts that a nonlinear dynamical system becomes less resilient to perturbations as it nears a bifurcation, or, `tipping point' \cite{Scheffer2009}. This phenomenon is expected to occur in systems which approach, either continuously \cite{Scheffer2009,Dakos2012} or incrementally \cite{Delecroix_CSD_Bursts,vanBeers2026_ews_for_loc}, a tipping point. In contrast, CSD cannot be expected for instantaneous transitions into instability (e.g., immediate and catastrophic faults). 

How CSD applies to control systems is intuitively understood through a linear system example: as a closed-loop system approaches instability - be this through changing system dynamics or controller behavior - the real part of its dominant eigenvalues approach zero, which directly results in a slowing system response. It is exactly this `slowing down' that CSD is attuned to for nonlinear control systems approaching instability. Consider a nonlinear (controlled) system of form eq. \ref{eq:nonlinearsys} where $x$ denotes the state vector and $u = k(x, r)$ provides the control law for tracking a reference signal, $r$.
\begin{equation}\label{eq:nonlinearsys}
    \dot{x} = f(x, k(x, r))
\end{equation}

At a bifurcation point, $x^{*}$, the Jacobian of the system, $J = \partial f/\partial x |x^{*}$, locally exchanges stability. Hence, as $x^{*}$ is approached, the real part of one of the eigenvalues of $J$ approaches zero which results in the slowing response rate of the (autonomous) controlled system to perturbations (i.e., inputs). What makes CSD so powerful in practice is that the implications of this eigenvalue movement can be observed without computing the (often unknown) system model and Jacobian. Instead, the slowing recovery rate can monitored through statistical indicators derived from system measurements \cite{Scheffer2009}. This makes CSD especially useful for monitoring the degradation in stability of nominally stable control systems. 

Conventionally, CSD is observed through the variance ($SD$) and lag-1 autocorrelation ($AC1$) computed along a sliding window applied over a (detrended) measurement signal of interest. As a tipping point is reached, it is expected that $AC1 \to 1$ and $SD \to \infty$ \cite{Scheffer2009}. Nonetheless, a core challenge persists in that there is little guidance on how to choose suitable parameters for these indicators \cite{vanBeers2026_ews_for_loc} and a poor selection can obscure an approaching critical transition \cite{Wen2018_MissingCSD}. 

Here, we provide a few guidelines that can help identify appropriate parameters for control systems using insights from control theory. These rules-of-thumb are constructed around the $AC1$ indicator, which is favored for its bounded and consistent behavior as a tipping point is approached \cite{Dakos2012}. Hence, the objective is to choose indicator parameters that help observe meaningful shifts in stability:

\begin{enumerate}
    \item \textbf{Measurement signal:} Closed-loop instabilities are likely to manifest in the controller actions \cite{vanBeers2026_ews_for_loc}. Consequently, the controller output(s) or actuator response(s) are good candidates for early warning of instability. Alternatively, other variables which are known to reflect the tipping point (e.g., inverted pendulum angle) may also be suitable candidate variables.
    \item \textbf{Detrending parameters:} Detrending is recommended to remove low-frequency trends in the nominal system behavior \cite{Scheffer2009,Dakos2012}. For control systems, this relates to the nominal control behavior. Moreover, the concern of worsening closed-loop stability often lies at higher frequencies where substantial system activity can provoke instability (e.g., through out-of-phase input-output behavior). These problematic high-frequency regions can be isolated from the nominal control behavior by choosing a cutoff frequency, $F_{m}$, to constrain the detrending. Without loss of generality, we opt for a trailing moving average detrender for runtime simplicity. The associated window size, $D$, can be obtained via:
    \begin{equation}\label{eq:ma_dtr__window}
        D \leq \frac{F_{s}}{4F_{m}}
    \end{equation}
    Where $F_{s}$ is the sampling rate. If there is no clear $F_{m}$, then choose $D$ such that the resultant detrended signal is a valid lag-1 autoregressive process \cite{vanBeers2026_ews_for_loc,Scheffer2009}.
    \item \textbf{Lag-1 autocorrelation (AC1) window:} The detrending step, in principle, has attenuated the nominal corrective control behavior below $F_{m}$. Thus, by choosing an $AC1$ window $W\geq8D$, frequencies $F \geq F_{m}$ can be fully observed within $W$. Large $W$ help observe a persistent change in resilience whereas shorter $W$ help recognize short - potentially temporary - excursions in resilience. 
\end{enumerate}

Following these parameterization steps effectively provides a proxy statistic for the system's recovery rate to high-frequency perturbations. This makes CSD particularly useful in a robustness context for detecting instabilities that emerge from poor high-frequency controller or system behavior. For these reasons, we hypothesize that simply observing the persistent corrective control behavior of a control system about a steady operating mode (e.g., hover for a quadrotor) already provides insight into its overall resilience. That is, if a concerning loss of stability is already evident at the steady state mode, then the system becomes especially vulnerable to instability at more demanding operational modes. 

Nonetheless, the inherent focus on high-frequency behavior can make it difficult to recognize emerging low-frequency instabilities that the nominal controller is eventually unable to stabilize. While the stabilizing effort may be reflected by changes in the controller's behavior (also at high-frequencies), choosing large $W$ can improve sensitivity to the low-frequency instabilities attenuated by the (moving average) detrender.

\section{Simulation studies}
Simulations of nonlinear control systems, an inverted pendulum on a cart and a quadrotor, are used to systematically drive them towards instability. Through these simulations, we show how the CSD indicators of resilience constructed following Section \ref{sec:csd} can be used to effectively monitor loss of stability.

\subsection{Inverted pendulum on a cart}
The inverted pendulum on a cart is a canonical nonlinear control system notorious for its unstable upright equilibrium. As such, the resilience of a closed-loop configuration can be measured based on how capable the controller is at maintaining the upright equilibrium. 

\subsubsection{Closed-loop dynamics}\label{sec:cartpoleModel}
The inverted pendulum on a cart system is represented by: 
\begin{equation}\label{eq:cartpole}
\begin{array}{c}
(M+m)\ddot x
+
ml\cos(\theta)\ddot\theta
=
u_{act}-b\dot x
+
ml\sin(\theta)\dot\theta^2
\\
ml\cos(\theta)\ddot x
+
\left(I+ml^2\right)\ddot\theta
=
-c\dot\theta
+
mlg\sin(\theta).
\end{array}
\end{equation}

\noindent where $x$ is the cart position and $\theta$ is the pole angle with $\theta = 0$ corresponding to the upright equilibrium. $M$ and $m$ are the cart and pole masses, respectively. The pole length is given by $l$ and its inertia by $I$. $g$ denotes the gravitational constant. The following nominal parameters are chosen: $M = 2.0$ kg, $m=0.5$ kg, $l=0.5$ m, $I = 0.005$ kgm$^{2}$, $c=0.01$ Nm$s$, $b=0.1$ N$s$m$^{-1}$, and $g=9.81$. 

The upright equilibrium is stabilized via nonlinear model predictive control (NMPC), with cost function:
\begin{equation}
    J = J_{K} +\sum_{k=1}^{K-1} \left ( J_{k} 
    + u_{k}^{T}Ru_{k} 
    + \Delta u_{k}^{T} S \Delta u_{k} \right )
\end{equation}
\noindent where:
\begin{equation*}
\begin{array}{c}
    J_{k} = (z_{k}-r_{k})^{T}Q(z_{k}-r_{k}) \\
    J_{K} = (z_{K}-r_{K})^{T}P(z_{K}-r_{K})
\end{array}
\end{equation*}
\noindent with $k$ denoting the current time step. $z_{k} = \left [ x, \dot{x}, \theta, \dot{\theta} \right]^{T}$, $r_{k}$, and $u_{k}$ represent the associated state, reference, and input, respectively. $\Delta u_{k} = u_{k}-u_{k-1}$ describes the difference between consecutive inputs. $Q$, $P$, $R$, and $S$ denote weight penalty matrices for the tracking error, terminal state, control effort, and control aggressiveness, respectively. The objective is to find the optimal control sequence $U^{*} = \text{arg}\min_{U}J(z_{k}, U)$ over a model prediction horizon, $N_{p}$, and control horizon $N_{c} < N_{p}$. The sequence of actions, $U^{*}$, is optimized over $N_{c}$ with the last action held constant until the end of the prediction horizon, $N_{p}$. Then, the first action in the sequence $U^{*}$ is taken and held constant over two samples for computational efficiency (i.e., the controller updates at half the rate of the simulation). Then optimization routine repeats from the new state. We elect $N_{p} = 0.8$ s, $N_{c} = 0.2$ s, and:
\begin{equation*}
    \begin{array}{cc}
        Q = \text{diag}(1.0, 20.0, 0.3, 2.0),
        &
        R = 0.01
        \\
        P = \text{diag}(50.0, 1000.0, 1.00, 100.0),
        &
        S = 0.002
    \end{array}
\end{equation*} 
Furthermore, to complicate the control task, the cart is driven by an actuator which supplies a force through eq. \ref{eq:u_act_cartpole} where $u_{act}$ is the true action and $u_{cmd}$ is the output of the NMPC, which is unaware of these actuator dynamics. $\tau = 33.3$ ms is the actuator time constant.  
\begin{equation}\label{eq:u_act_cartpole}
   \dot{u}_{act} = \frac{-1}{\tau}u_{act} + \frac{1}{\tau}u_{cmd}
\end{equation}
Simulations are conducted at 100 Hz with the state and actuator dynamics integrated via a 4th order Runge-Kutta scheme. Likewise, the NMPC internal model updates during optimization are also integrated the same way. Note that since every other control action is held constant, the controller's effective update rate is 50 Hz and thus $N_{p}=40$ and $N_{c}=10$ samples. White measurement noise is injected into the state vector at each time step, with scales: $\sigma_{x}=0.01$ m, $\sigma_{\theta} = 0.01$ rad, $\sigma_{\dot{x}} = 0.001$ m$s^{-1}$, and $\sigma_{\dot{\theta}}=0.001$ rad$s^{-1}$.

\subsubsection{Indicators of resilience}

Following the procedure in Section \ref{sec:csd}, the actuator response measurement, $u_{act}$, is used as the basis of the resilience indicator. Around the upright equilibrium, a well-behaved controller should have a closed-loop bandwidth substantially below the control update rate of $50$ Hz. Likewise, the actuator time constant is $\tau = 33.3$ ms, leading to a corner frequency of around $5$ Hz. As such we choose a moving average window of $D=4$ samples (corresponding to $F_{m} = 6.25$ Hz) to remove much of the low frequency content that represents the nominal controller behavior. Subsequently, the AC1 is sized as $W=8\cdot4=32$ samples. This early warning signal design is kept consistent across all following simulations of the cart pendulum system. 

\subsubsection{Closed-loop instability}\label{subsubsec:cartpole_destab}

\begin{table*}
    \caption{Resilience metrics for the simulation experiments using the inverted pendulum on a cart system.}
    \label{tab:cartpole}
    \centering
    \begin{tabular}{cccccccc}
         \textbf{Experiment} & $\mathbf{M}$ (kg) & $\mathbf{d}$ & \textbf{Actuator} $\boldsymbol{\tau}$ (ms) & \textbf{Percentage escaped$^{*}$} & \textbf{Mean escape time} (s) & \textbf{MAE$^{**}$ in} $\boldsymbol{\theta}$ (rad) & \textbf{Median AC1}$^{\dagger}$\\ \hline
         Section \ref{subsubsec:cartpole_destab} & 2.0 & 0.5 & 33.3 & 0 & - & 2.90$\times 10^{-3}$ & 0.73 \\
         & 2.0 & 1.0 & 33.3 & 0 & - &  2.36$\times 10^{-3}$ & 0.74 \\
         & 2.0 & 2.0 & 33.3 & 0 & - &  2.81$\times 10^{-3}$ & 0.79 \\
         & 2.0 & 2.5 & 33.3 & 0 & - &  1.03$\times 10^{-2}$ & 0.83 \\
         & 2.0 & 3.0 & 33.3 & 98 & 4.42 &  0.544 & - \\
         Section \ref{subsubsec:cartpole_pert} & 0.1 & 1.0 & - & 100 & 0.77 & 2.723 & - \\
         & 0.1 & 1.0 & 33.3 & 100 & 0.59 & 2.821 & - \\
         & 0.20 & 1.0 & 33.3 & 1 & 3.10 & 1.84$\times 10^{-2}$ & 0.79 \\
         & 1.00 & 1.0 & 33.3 & 0 & - & 2.86$\times 10^{-3}$ & 0.75 \\
         & 2.00 & 1.0 & 33.3 & 0 & - & 2.75$\times 10^{-3}$ & 0.71 \\
         & 3.00 & 1.0 & 33.3 & 0 & - & 2.75$\times 10^{-3}$ & 0.70 \\
         
         \hline
    \end{tabular}
\vspace{0.5mm}
\parbox{0.85\textwidth}{
\scriptsize
    $^{*}$ The pendulum is considered to have diverged, or `escaped', from the upright equilibrium when $|\theta|\geq 90$ deg, indicating that it has `hit' the ground.
    
    $^{**}$ MAE stands for mean absolute error.
    
    $^{\dagger}$ AC1 values for simulations that diverged are not shown as the tipping point has passed; they contain system behavior in the post-tipping regime. 
}
\end{table*}

This first experiment serves to demonstrate that CSD indeed reflects an unambiguous deterioration in nonlinear system stability. To achieve this, the system dynamics in eq. \ref{eq:cartpole} are augmented with a synthetic destabilizing amplifier, $d$, that modulates the natural destabilizing dynamics of the cart pole system (i.e., the $mlg\sin(\theta)$ term) without altering any of its other properties:
\begin{equation*}
    \begin{array}{c}
(M+m)\ddot x
+
ml\cos(\theta)\ddot\theta
=
u_{act}-b\dot x
+
ml\sin(\theta)\dot\theta^2
\\
ml\cos(\theta)\ddot x
+
\left(I+ml^2\right)\ddot\theta
=
-c\dot\theta
+
mlg\sin(\theta)d.
\end{array}
\end{equation*}
The cart pole system is progressively brought closer to closed-loop instability by increasing the destabilizing factor $d\in\{0.5, 1.00, 1.50, 2.00, 2.50, 3.00\}$. For each $d$, 100 unique seed noisy simulations are initialized with $z_{0} = [0.1, 0, 0, 0]$. Note that a small initial position perturbation is provided to observe the destabilizing effect since, near the upright equilibrium, it is almost negligible through $\sin(\theta)\approx0$. Across all simulations, the cart properties and NMPC construction are kept consistent and follow from Section \ref{sec:cartpoleModel}. Likewise, the internal model of the NMPC perfectly matches the true system, though it does not contain the actuator dynamics. 

\begin{figure}
    \centering
    \includegraphics[width=\linewidth]{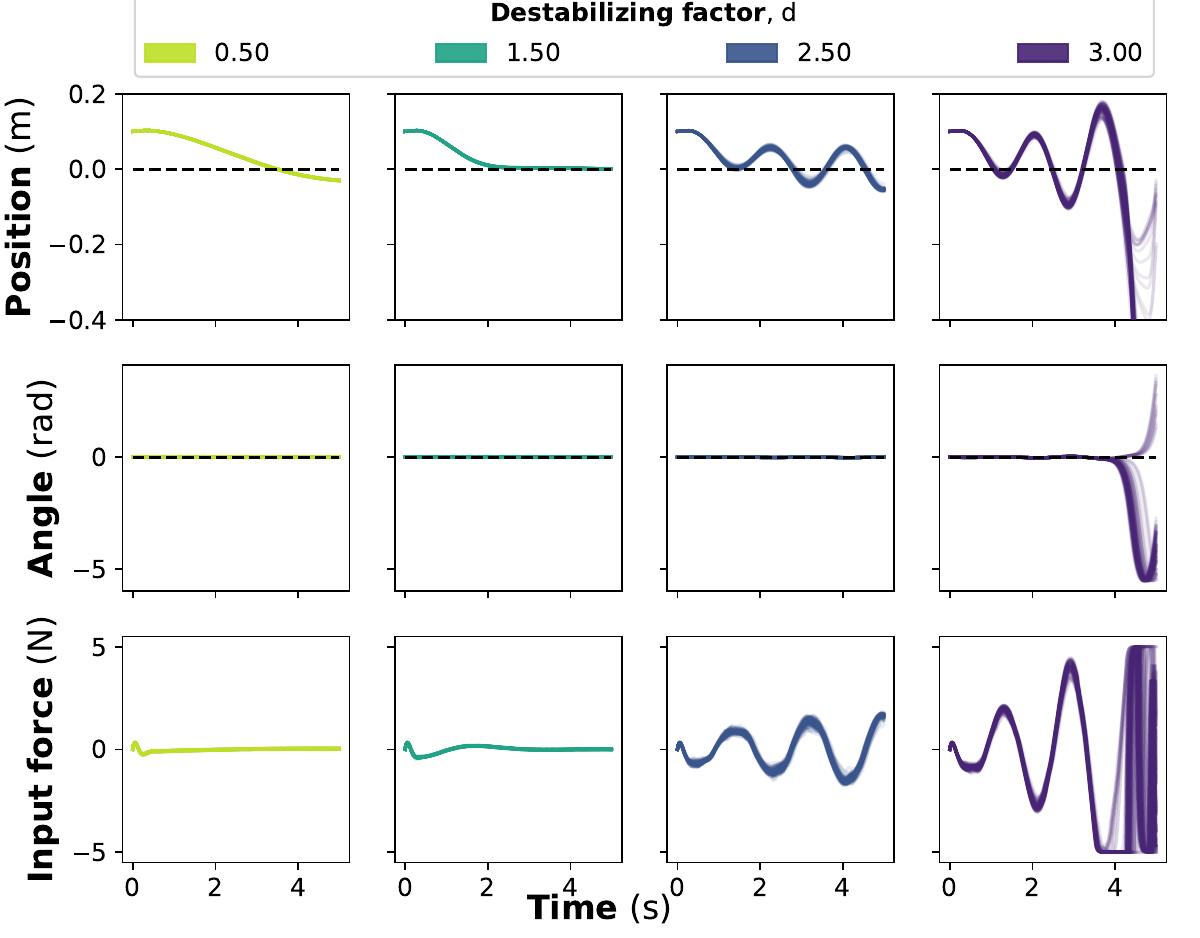}
    \caption{Simulated state and input evolution of the inverted pendulum on a cart under different destabilizing amplification factors, $d$.}
    \label{fig:cart_pole__destab_state}
\end{figure}

\begin{figure}
    \centering
    \includegraphics[width=\linewidth]{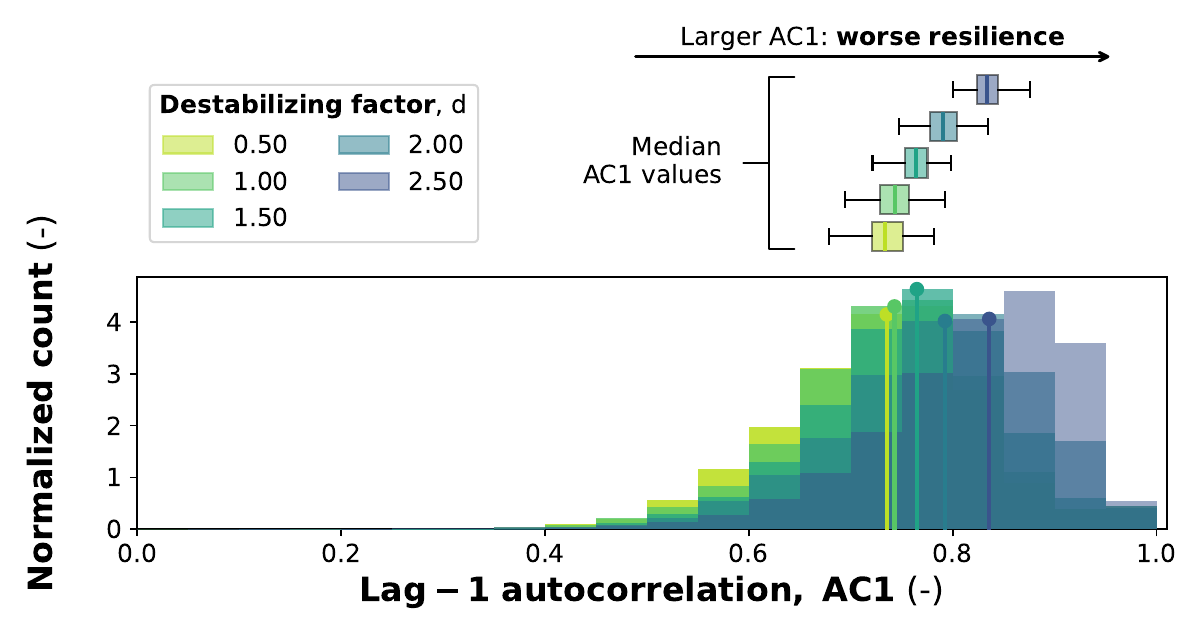}
    \caption{Indicators of resilience, AC1, observed for the inverted pendulum on a cart simulations subject to varying destabilizing amplification factors with $d=1.00$ representing the nominal system.}
    \label{fig:cart_pole__destab_csd}
\end{figure}

As $d$ increases, the destabilizing torque intensifies and weakens the controller's capacity to maintain the upright equilibrium. This is apparent for $d\geq2.5$ where an unstable growth in input response emerges, shown in fig. \ref{fig:cart_pole__destab_state}. Moreover, in 98\% of the simulations with $d=3.0$, the closed-loop system diverges completely from the upright equilibrium with an average escape time (i.e., time until the pendulum `hits' the ground at $|\theta| = 90$ deg) of $t=4.42$ s, summarized in table \ref{tab:cartpole}. This approach towards instability (i.e., for $d<2.5$) is not immediately evident from the error statistics (table \ref{tab:cartpole}) or system response alone. In contrast, this degradation in closed-loop stability as $d$ grows is apparent early on through the indicators of resilience, shown in fig. \ref{fig:cart_pole__destab_csd} and table \ref{tab:cartpole}.  

\subsubsection{Closed-loop resilience to perturbations}\label{subsubsec:cartpole_pert}
This experiment explores how CSD works also as an indicator of nonlinear control system resilience to finite perturbations (i.e., stability margin). Here, the stability margin of the cart pendulum system is reduced by increasing the closed-loop system bandwidth beyond the capabilities of the controller and actuators. This is achieved by decreasing the cart mass, $M = 2 \to M = 0.1$, for both the true system and NMPC internal model. As the cart mass decreases, the open-loop unstable upright pole grows in magnitude inversely proportional to the cart mass. The NMPC internal model recognizes this and thus seeks to react faster to stabilize the upright equilibrium. 

Such behavior becomes problematic when the closed-loop system bandwidth approaches the control loop frequency where, eventually, the ability of the controller to regulate the system becomes increasingly sluggish (i.e., slow) with respect to the underlying dynamics. This renders the system vulnerable to self-reinforcing feedback loops (i.e., unstable behavior) that emerge when measurement noise is introduced into the system. The persistent response of the controller to this (high-frequency) noise excites the fast system dynamics that it is now too slow to adequately regulate. 

\begin{figure}
    \centering
    \includegraphics[width=\linewidth]{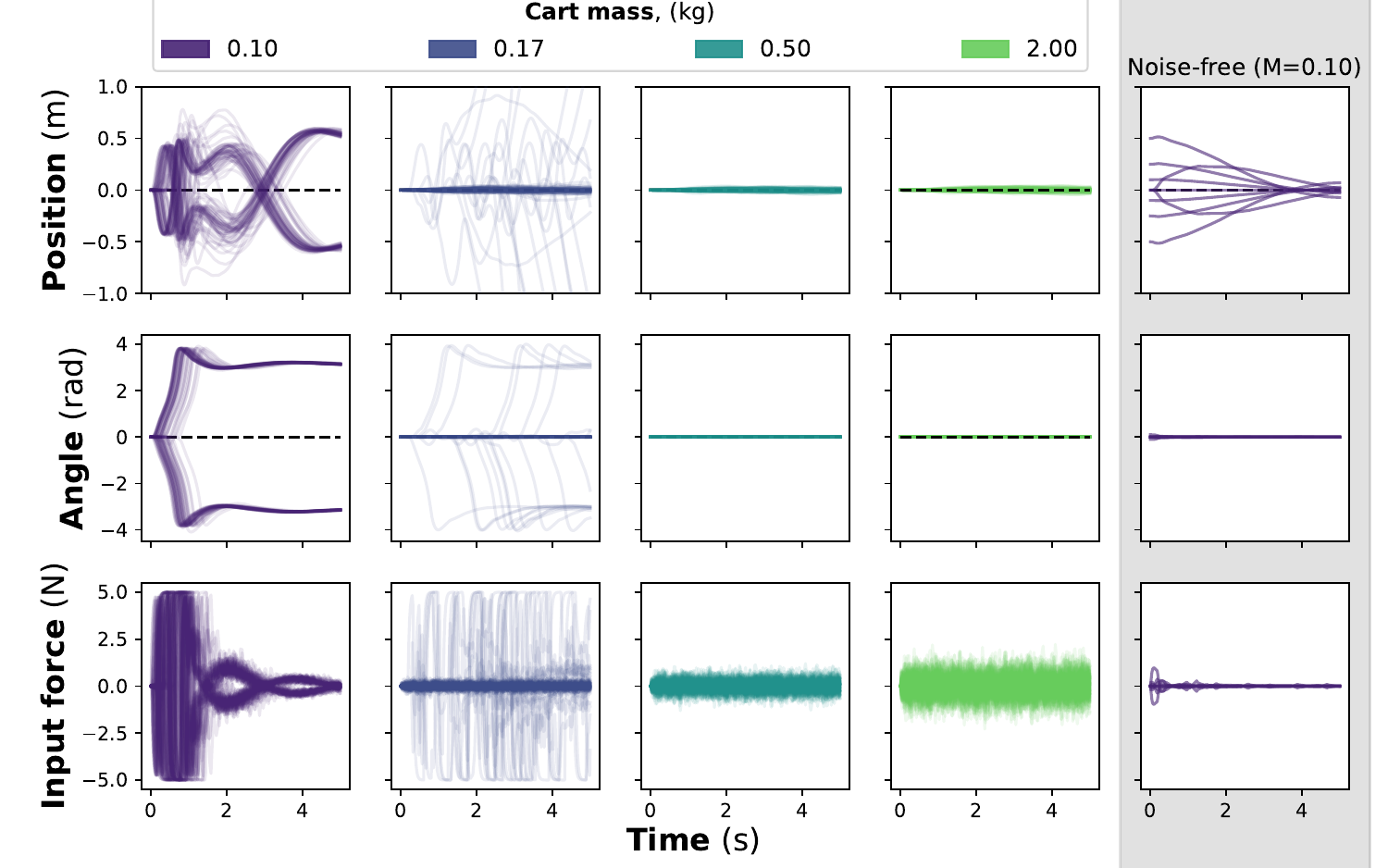}
    \caption{State and input evolution of the actuated ($\tau = 33.3$ ms) inverted pendulum on a cart, as a function of the cart mass ($M$), for different noisy simulation runs. The right-most plots depict the noise-free stabilization of the upright equilibrium with $M=0.1$.}
    \label{fig:cart_pole__perturb_state}
\end{figure}

\begin{figure}
    \centering
    \includegraphics[width=\linewidth]{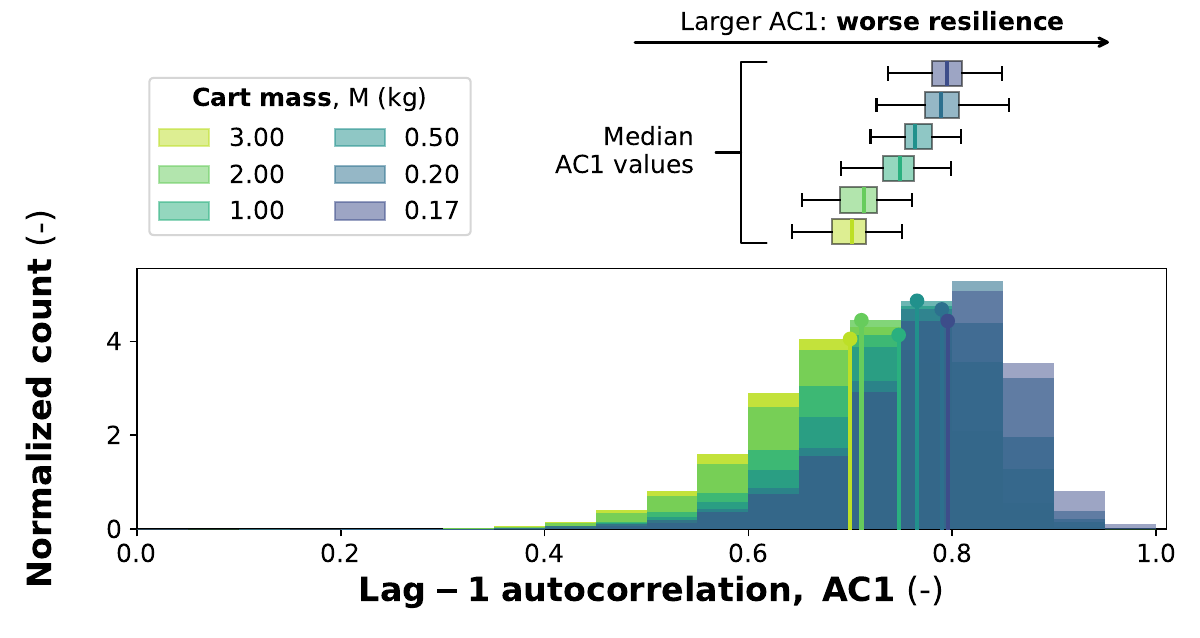}
    \caption{Indicators of resilience, AC1, observed for the noisy simulations of the actuated ($\tau = 33.3$ ms) inverted pendulum on a cart system under different cart masses.}
    \label{fig:cart_pole__perturb_csd}
\end{figure}

This noise-induced instability can be illustrated through simulations of the cart pendulum system. As a baseline, the right-most plots in fig. \ref{fig:cart_pole__perturb_state} illustrate how the cart pendulum system with $M=0.1$ can maintain stability without noise, including for states initialized outside the upright equilibrium and with unknown actuator dynamics. This is no longer possible when measurement noise is introduced: table \ref{tab:cartpole} summarizes the stochastic divergence characteristics for 100 unique-seed noisy simulations initialized at the upright equilibrium (i.e., $z_{0} = [0, 0, 0, 0]$) which also show that divergences occur even for simulations without actuator dynamics with an average escape time of 0.77 s. Thus, while the upright equilibrium remains locally stable in the noise-free scenario, the closed-loop system has become less resilient to finite perturbations which now more easily induce instability as $M$ decreases, shown in fig. \ref{fig:cart_pole__perturb_state} and table \ref{tab:cartpole}. This decline in system resilience as the cart mass is decreased is evident through the CSD indicators well before the divergent behavior emerges, shown in fig. \ref{fig:cart_pole__perturb_csd}, and is already apparent at the (stabilized) upright equilibrium operating condition. 

We note that the NMPC used here is unaware of the actuator or noise (i.e., robust NMPC is not used). While such a design may certainly be used to improve controller robustness, that postpones the issue of instability studied here to stronger noise regimes. Instead, our objective is to demonstrate how CSD can be used to monitor for a deterioration in resilience of an arbitrary closed-loop system en-route to instability, which can also extend to more `robust' controller designs.

\subsection{Quadrotor}\label{sec:quadsim}
The application of CSD towards more conventional high-dimensional robotic systems is explored through simulations of a quadrotor. This system is chosen as recent work has shown that CSD reflects quadrotor loss of stability in real world flight tests \cite{vanBeers2026_ews_for_loc}, yet it remains unclear whether this relation between CSD and quadrotor stability is more fundamental. Here, we show that CSD also emerges in simulated models of a quadrotor suffering loss of thrust effectiveness, lending support to such theoretical connections. Moreover, we show that loss of stability can already be observed through hovering flight, without needing to expose the quadrotor to the more demanding maneuvers that tend to result in instability.

\subsubsection{Closed-loop dynamics}
A rotor local model, given by eq. \ref{eq:appE_rtrlcl_model_F} and \ref{eq:appE_rtrlcl_model_M}, is used to facilitate the quadrotor fault simulations with loss of effectiveness (LOE) in individual rotor thrust. 
\begin{equation}\label{eq:appE_rtrlcl_model_F}
    \dot{V}
    =
    R\left[ 
    \begin{array}{c}
         0 \\ 0 \\ g
    \end{array}
    \right ]
    -
    \Omega \times 
    V
    +
    \frac{1}{m}\left [  
    \begin{array}{c}
         0 \\ 0 \\ -\sum_{i=1}^{4}\kappa_{i}\omega_{i}^{2} 
    \end{array}
    \right ]
\end{equation}
\begin{equation}\label{eq:appE_rtrlcl_model_M}
    \dot{\Omega}
    = 
    - I_{v}^{-1} \left (
    \Omega
    \times \left (
    I_{v} \Omega
    \right )
    \right )
    + I_{v}^{-1} M_{c} \boldsymbol{\omega}^{2}
\end{equation}

\noindent where $V$ represents the body translational velocity, $\Omega$ the body rotational rates, and $\boldsymbol{\omega}=[\omega_{1}, \omega_{2}, \omega_{3}, \omega_{4}]^{T}$ defines the rotor speed inputs. The quadrotor mass and moment of inertia are given by $m$ and $I_{v}$, respectively. $g$ represents the gravitational constant, which is transformed into the quadrotor body frame (from the inertial north-east-down reference frame) via the rotation matrix $R$. The torque control effectiveness matrix, $M_{c}$, is given by:
\begin{equation}
    M_{c} =\left [ 
    \begin{array}{cccc}
         -b\kappa_{1} & -b\kappa_{2} & b\kappa_{3} & b\kappa_{4} \\
         -b\kappa_{1} & b\kappa_{2} & -b\kappa_{3} & b\kappa_{4} \\
         \nu_{1} & -\nu_{2} & -\nu_{3} & \nu_{4}
    \end{array}
    \right ] 
\end{equation}
The thrust coefficient of the $i^{th}$ rotor is represented by $\kappa_{i}$ and $\nu_{i}$ denotes the yaw torque coefficient. $b$ is the moment arm of between the rotor center of thrust and the vehicle center of mass. The motors are driven by a first order actuator model where $\omega_{cmd,i}$ is the command provided to the $i^{\text{th}}$ rotor:
\begin{equation}
   \dot{\omega}_{i} = \frac{-1}{\tau_{i}}\omega_{i} + \frac{1}{\tau_{i}}\omega_{cmd,i}
\end{equation}

The quadrotor simulations rely on incremental nonlinear dynamic inversion (INDI) to facilitate flight control, adopting the architecture of \cite{SMEUR201879_INDI}: the position, velocity, and attitude of the quadrotor are controlled via a cascaded series of proportional derivative controllers while the inner rate and thrust loop is handled by INDI. We choose INDI for its disturbance rejection characteristics, which can also leave it vulnerable to singular perturbations \cite{TIJMEN2023_INDI_robustness}. 

\subsubsection{Simulation overview}
All quadrotor simulations are conducted in Python (3.13.5) and run at 250 Hz. The quadrotor model parameters are: $m=0.433$ kg, $g=9.81$ m$s^{-2}$, $I_{v}=\text{diag}(8.65\times 10^{-4}, 1.07\times 10^{-3}, 1.71\times 10^{-3})$ kgm$^{2}$, $b=0.077$ m, $\kappa_{i} = 8.72\times10^{-7}$ N$s^{2}$, $\nu_{i} = 4.29\times10^{-9}$ N$s^{2}$m, and $\tau_{i} = 17$ ms. Simulations are conducted with measurement (white) noise injected into the system states with scales: $\sigma_{att} = 0.01$ rad in attitude, $\sigma_{vel} = 0.01$ m$s^{-1}$ in velocity, and $\sigma_{rate} = 0.009$ rad$s^{-1}$ in body angular rates. Furthermore, the controller provides actions at each time step and quadrotor state derivatives are computed following eq. \ref{eq:appE_rtrlcl_model_F} and \ref{eq:appE_rtrlcl_model_M}. These state derivatives are then integrated using a 4th order Runge-Kutta scheme.

\subsubsection{Indicators of resilience}
Following \ref{sec:csd}, the rotor speed measurements, $\boldsymbol{\omega}$, are used to derive the indicators of resilience. As the controller updates at 250 Hz, it tracks input commands of below $\approx$ 25 Hz well. However, the motor actuator time constant for all rotors is $\tau_{1, 2, 3, 4}=17$ ms ($\approx 60$ s$^{-1}$), meaning that its ability to track signals above 10 Hz deteriorates. Considering both of these, a moving average detrending window of $D=3$ samples ($F_{m} \approx 20$ Hz) is chosen. Following \ref{sec:csd}, an AC1 window of $W=8\cdot D = 24$ samples is used.  

\subsubsection{Instabilities from system faults}
The quadrotor is brought closer to instability by increasing the LOE in per-rotor thrust, modeled by scaling the nominal thrust coefficient of an affected rotor, $\kappa_{i}$, following:
\begin{equation}\label{eq:loe}
\kappa_{dmg,i} = \lambda \kappa_{i}, \quad \lambda \in (0, 1]
\end{equation}

In this simulation study, the LOE affects only a single rotor at a time (i.e., no simultaneous LOE faults are conducted). While it is clear that, below a critical LOE threshold, the quadrotor cannot simultaneously maintain hovering flight with fixed attitude, the issues of instability can emerge earlier under more demanding flight maneuvers. As such, we explore the capacity of the quadrotor to follow three different trajectories - hovering flight, lemniscate of Gerono of amplitude 5 m at 5 m$s^{-1}$, and an ascending spiral of radius 3 m at 4 m$s^{-1}$ - under varying LOE severity. We consider the quadrotor to be in a state of `loss of control' when either the state trajectory diverges considerably from the reference (i.e., by position errors greater than 30 meters) or when the roll/pitch attitude of the quadrotor exceeds 90 degrees, which violates assumptions made regarding the permissible attitude in the controller design. For each trajectory tracking task, a set of 100 noisy simulations are conducted for each $\lambda \in \{1.0, 0.8, 0.5, 0.2, 0.1\}$, corresponding to LOE of $\{0, 20, 50, 80, 90\}$\% in thrust. 

\begin{figure}
    \centering
    \includegraphics[width=\linewidth]{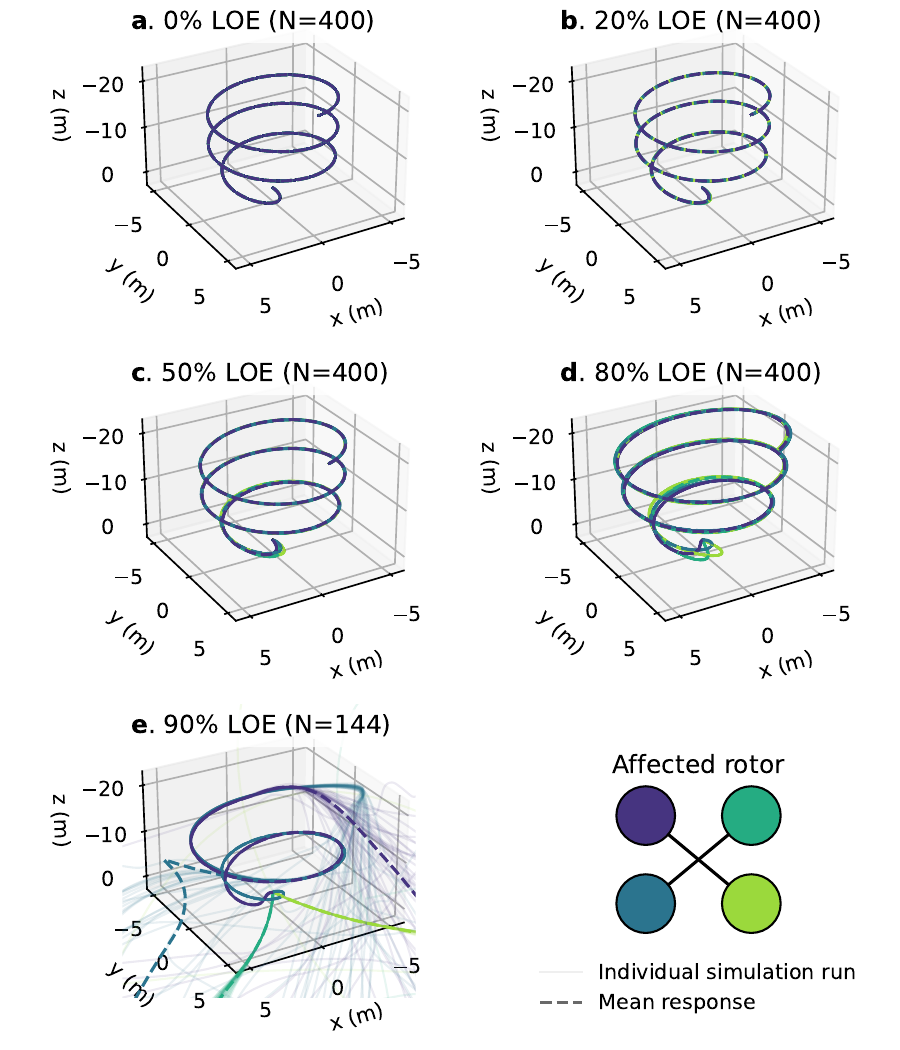}
    \caption{Simulated trajectories of the quadrotor following a spiral trajectory under various loss of effectiveness (LOE) in per-rotor thrust. N denotes the total number of flights completed (of 400) without numerical simulation divergence.}
    \label{fig:quad_sim__spirals}
\end{figure}

\begin{figure}
    \centering
    \includegraphics[width=\linewidth]{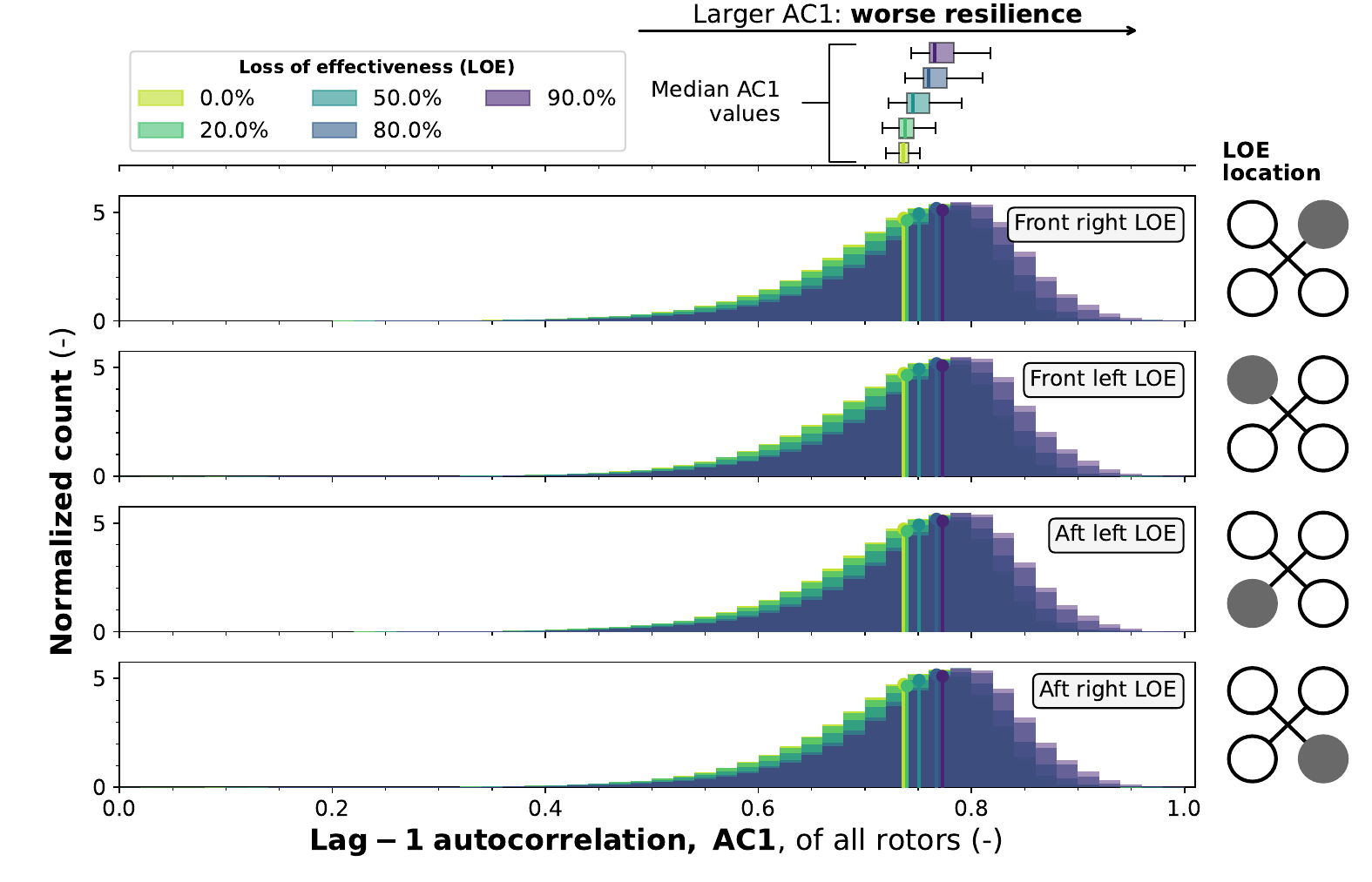}
    \caption{Indicators of resilience, $AC1$, observed during simulated hovering flights of the quadrotor under varying loss of effectiveness (LOE) severity levels in per-rotor thrust.}
    \label{fig:quad_sim__csd}
\end{figure}

At the hovering condition, the quadrotor is able to maintain stability across all LOE severity levels. However, the deterioration in closed-loop resilience becomes evident under the more demanding flight maneuvers: at an 80\% LOE severity, the quadrotor loses control in all of the leminscate tracking tasks. Likewise, at a 90\% LOE severity, the quadrotor loses control in all of the leminscate and spiral tracking tasks (see fig. \ref{fig:quad_sim__spirals}). Nonetheless, this proximity to instability of the high LOE conditions is already apparent at the hovering condition through an increase in the CSD indicators alongside LOE severity, shown in fig. \ref{fig:quad_sim__csd}. These results demonstrate that CSD can be used to identify abnormal system behavior even at relaxed operating modes.

\section{Real-world experiments}

\begin{figure*}
    \centering
    \includegraphics[width=\linewidth]{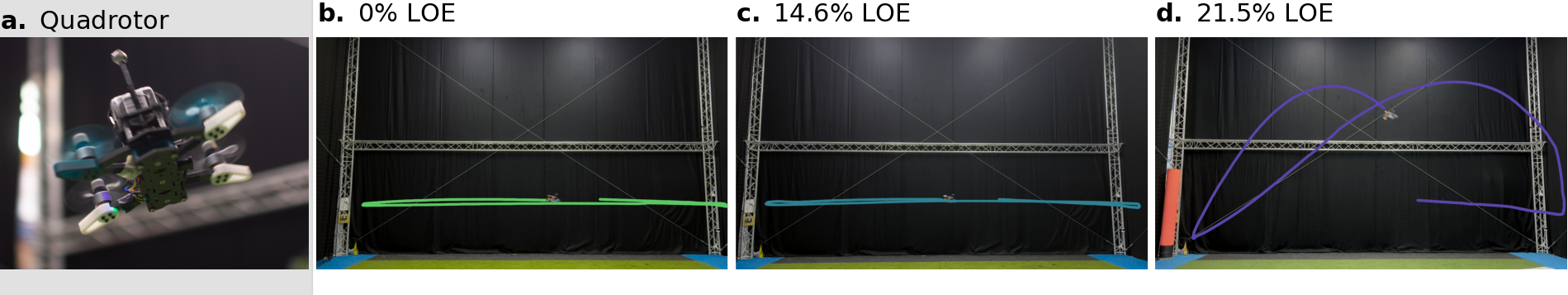}
    \caption{Quadrotor flight experiments. \textbf{a} depicts the quadrotor platform used to gather data and \textbf{b}-\textbf{d} depict portions of the leminscate of Gerono trajectories under different loss of effectiveness (LOE) severity levels. Frequent flyaway events occurred for the 21.5\% LOE case, shown in \textbf{d}.}
    \label{fig:quad_exp}
\end{figure*}

We replicate the quadrotor simulation study by conducting flight experiments using a 3-inch quadrotor (fig. \ref{fig:quad_exp} \textbf{a}) with portions of it's propeller blades cut off (see fig. \ref{fig:quad_exp__leminscate}). The resultant LOE is estimated via the proportional change in rotor speed of the affected motor required to sustain hovering flight. Overall, the quadrotor is expected to be less resilient to LOE compared to the simulation studies due to the vibrations that emanate from the damaged blades, which are unmodeled in the simulator. These additional vibrations can amplify the destabilizing mechanisms in the closed-loop system, similar to those seen for the inverted pendulum in Section \ref{subsubsec:cartpole_pert}. 

In these experiments, autonomous flight is facilitated by Indiflight\footnote{Available here: \href{https://github.com/tudelft/indiflight}{https://github.com/tudelft/indiflight}} (an INDI fork of Betaflight). While an external motion capture system is used to measure the quadrotor's position, it only facilitates trajectory tracking and is not needed for the indicators of resilience derived here. Flight data is recorded at 1000 Hz and is resampled to a constant 500 Hz. 

\subsection{Indicators of resilience}
The indicator processing parameters are chosen for consistency with those of the simulation study in Section \ref{sec:quadsim}. To this end, the rotor speed measurements, $\boldsymbol{\omega}$, again form the basis of the resilience indicators. These measurements are made available by the onboard ESC through Bidirectional DShot. Moreover, we select $D=6$ such that $F_{m} \approx 20$ Hz. Then, the AC1 window is $W = 8\cdot D = 48$ samples.

\subsection{Quadrotor flyaway events}

\begin{figure}
    \centering
    \includegraphics[width=\linewidth]{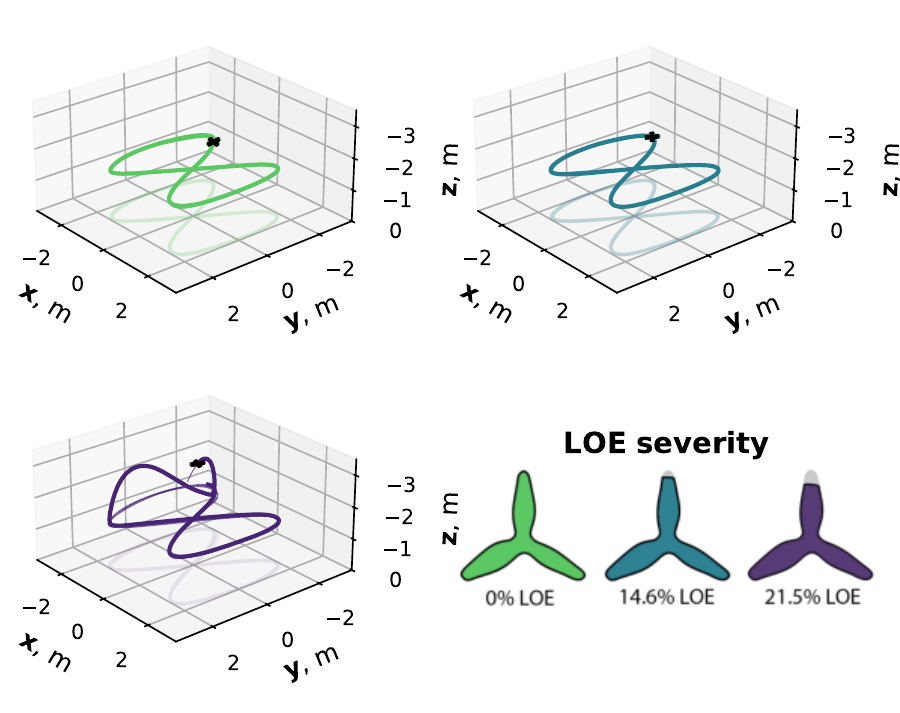}
    \caption{Re-visualization of the real-world lemniscate of Gerono trajectories (fig. \ref{fig:quad_exp}) flown under different loss of thrust effectiveness (LOE) severity levels.}
    \label{fig:quad_exp__leminscate}
\end{figure}

\begin{figure}
    \centering
    \includegraphics[width=\linewidth]{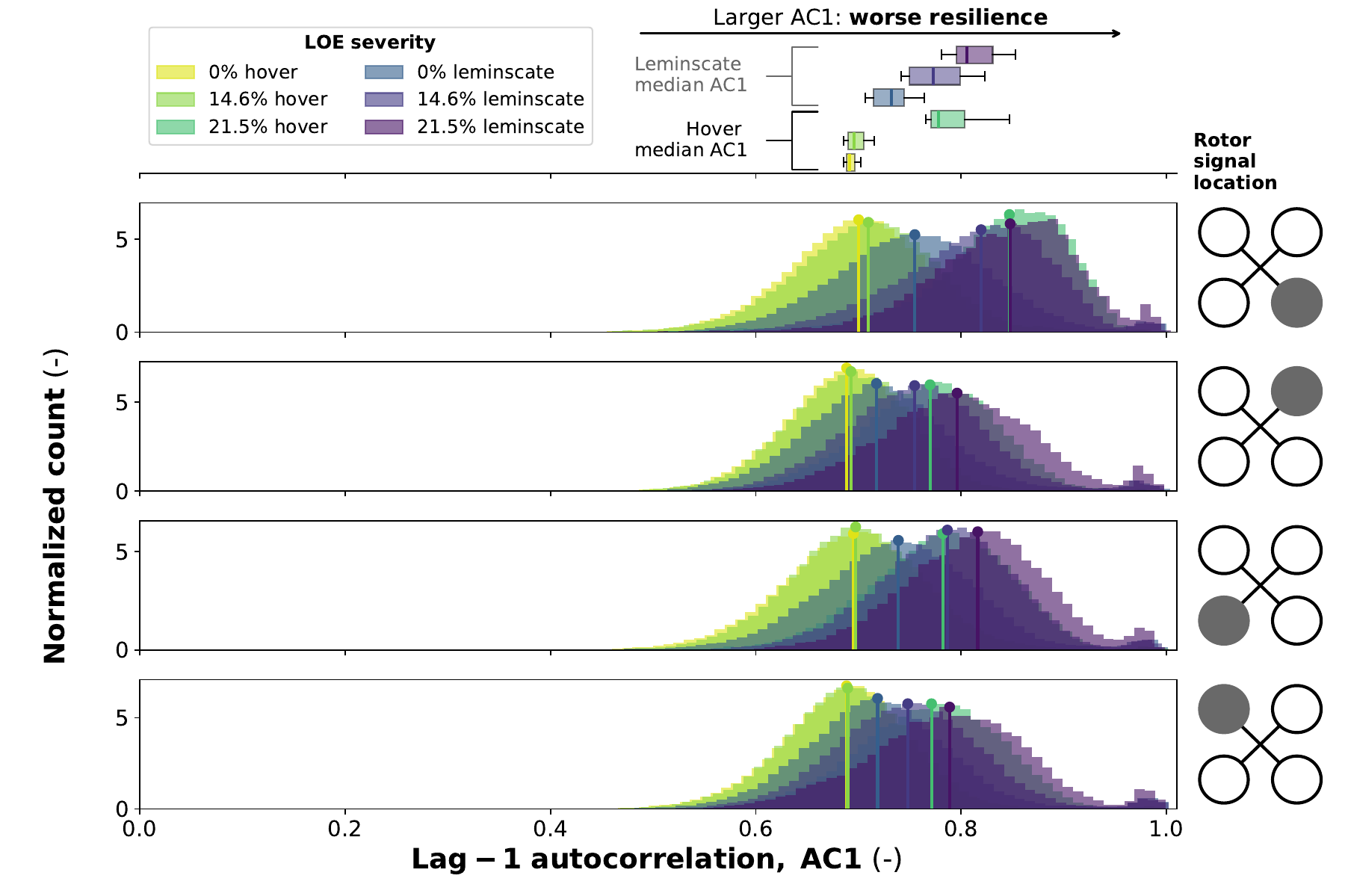}
    \caption{Indicators of resilience, $AC1$, observed during real world flights of the quadrotor under different loss of effectiveness (LOE) severity levels and trajectory tasks.}
    \label{fig:quad_exp__csd}
\end{figure}

As a baseline, five individual hovering flights are flown under three different LOE $\in \{0\%, 14.6\%, 21.5\%\}$ conditions experienced by the aft left rotor. Figure \ref{fig:quad_exp__csd} depicts the associated median AC1 values, which indicate that the 21.5\% LOE is significantly less resilient than the nominal (i.e., 0\%) and 14.5\% LOE conditions. Interestingly, the hovering CSD results indicate no apparent difference in resilience between the nominal and 14.5\% LOE conditions. This highlights one of the limitations of the locality inherent to CSD: it reports how the system \textit{locally} relates to instability. 

Nonetheless, the degradation in system resilience provoked by the blade damage emerges under more demanding tracking tasks. At each LOE severity, five flights are conducted where the quadrotor follows a leminscate of Gerono trajectory of amplitude 2 m at 2.5 m$s^{-1}$. While the quadrotor suffering 14.5\% LOE manages to effectively follow the desired trajectory (fig. \ref{fig:quad_exp} and \ref{fig:quad_exp__leminscate}), the associated increase in AC1 values over the healthy rotor case indicates that the system now operates closer to instability (fig. \ref{fig:quad_exp__csd}). Indeed, at the 21.5\% LOE severity, the quadrotor often loses control - three of the five leminscate flights resulted in a crash whereas the remaining two involved flyaways (i.e., uncontrolled ascending flight) throughout the tracking task, seen in fig. \ref{fig:quad_exp} \textbf{d} and fig. \ref{fig:quad_exp__leminscate}. These results show how CSD is also a function of the operating regime of the system: more aggressive operation can render the system more vulnerable to instability. Nonetheless, whether such operation is a concern for stability can already be observed at more relaxed operating conditions through the resilience indicators.

\section{Conclusion}
Our results show how critical slowing down (CSD) can reflect the (local) proximity of a nonlinear control system to instability. In the context of nominally stable autonomous control systems, CSD provides a powerful tool to monitor the degradation in closed-loop stability margin solely from online measurement data and without the need for system models. Using insights from control theory, we provide a systematic guideline for how one may construct such generic indicators of resilience for an arbitrary nonlinear control system. 

The eventual goal of promoting safety in autonomous and robotic systems is to prevent the occurrence of instability. As such, future work should investigate ways of applying these CSD-based indicators of resilience towards preventing instability or to promote safe learning. For example, these indicators can be used to inform a controller of what actions work to deteriorate resilience and which actions remain safe. 

\section*{Acknowledgments}

This work is supported through the European Research Council Consolidator Grant 101126132 on ``Autonomous robots with common sense" (ARCS). 

\bibliographystyle{IEEEtran}
\bibliography{ref}

\newpage

\vfill

\end{document}